\documentclass[%
 reprint,
showpacs,
 amsmath,amssymb,
 aps,
]{revtex4-1}

\usepackage{graphicx}
\usepackage{dcolumn}
\usepackage{bm}

\usepackage{anysize}
\marginsize{1.75cm}{1.2cm}{0.3cm}{3.cm}

\newcommand{\be}{\begin{equation}}
\newcommand{\ee}{\end{equation}}
\newcommand{\bea}{\begin{eqnarray}}
\newcommand{\eea}{\end{eqnarray}}

\newcommand{\nfj}{N_F^j}
\newcommand{\bM}{\bar{M}}
\newcommand{\tm}{\tilde{M}}

\usepackage{color} 
\newcommand{\yon}[1]{}
\newcommand{\gui}[1]{}

\begin{document}

\title{Fracture in Disordered Heterogeneous Materials as a Stochastic Process}

\author{Yon Visell$^{1,}$}
 \email{yon.visell@drexel.edu}
\author{Guillaume Millet$^2$}%
\affiliation{%
$^1$Drexel University, Electrical and Computer Engineering Dept., Philadelphia, USA \\
$^2$Sorbonne Universit\'es, UPMC Univ.~Paris 06, ISIR, Paris, France 
}


\date{\today}

\begin{abstract}
Fracture processes in heterogeneous materials comprise a large number of disordered spatial degrees of freedom, representing the dynamical state of a sample over the entire domain of interest. This complexity is usually modeled directly, obscuring the underlying physics, which can often be characterized by a small number of physical parameters. In this paper, we derive a closed-form expression for a low dimensional model that reproduces the stochastic dynamical evolution of time-dependent failure in heterogeneous materials, and efficiently captures the spatial fluctuations and critical behavior near failure. Our construction is based on a novel time domain formulation of Fiber Bundle Models, which represent spatial variations in material strength via lattices of brittle, viscoelastic fiber elements. We apply the inverse transform method of random number sampling in order to construct an exact stochastic jump process for the failure sequence in a material with arbitrary strength distributions. We also complement this with a mean field approximation that captures the coupled constitutive dynamics, and validate both with numerical simulations. Our method provides a compact representation of random fiber lattices with arbitrary failure distributions, even in the presence of rapid loading and nontrivial fiber dynamics.

\end{abstract}

\pacs{46.50.+a,62.20.M-,64.60.-i}
\maketitle


Failure processes in disordered, heterogeneous materials (e.g., fiberglass, wood, asphalt)
have attracted interest in scientific and engineering research because of the complexity of
the phenomena they exhibit. 
A full microscopic understanding of structural failure in such materials remains elusive, due to their disordered nature and large number of constituent elements.  Nonetheless, aspects of these processes can be captured via comparatively simple statistical models \cite{alava2006statistical,herrmann1990statistical}.  While fracture evolution is guided by the complex spatial composition of the material, the pattern of  temporal failures involved can also be considered to richly encode this spatial disorder.  Generative models of temporal failure in such materials typically require that the state of a large number of spatial degrees of freedom be updated.  

Our main contribution is an explicit temporal model of fracture that captures the stochastic temporal dynamics without representing spatial degrees of freedom.
We identify a generative expression for a stochastic jump process exactly capturing the fluctuating pattern of failure in a Fiber Bundle Model of fracture, and further obtain a novel factorization, separating out a mean constitutive response that closely matches the averaged nonlinear stress-strain behavior of the exact model.  The latter aspect allows the (smooth) stress-strain dynamics to be simulated deterministically, without tracking  the (rapidly fluctuating) random failure history in the material.  The former provides an iterative stochastic description of instants of failure in the material as it is loaded.

\begin{figure}
\begin{center}
\includegraphics[width=7.75cm]{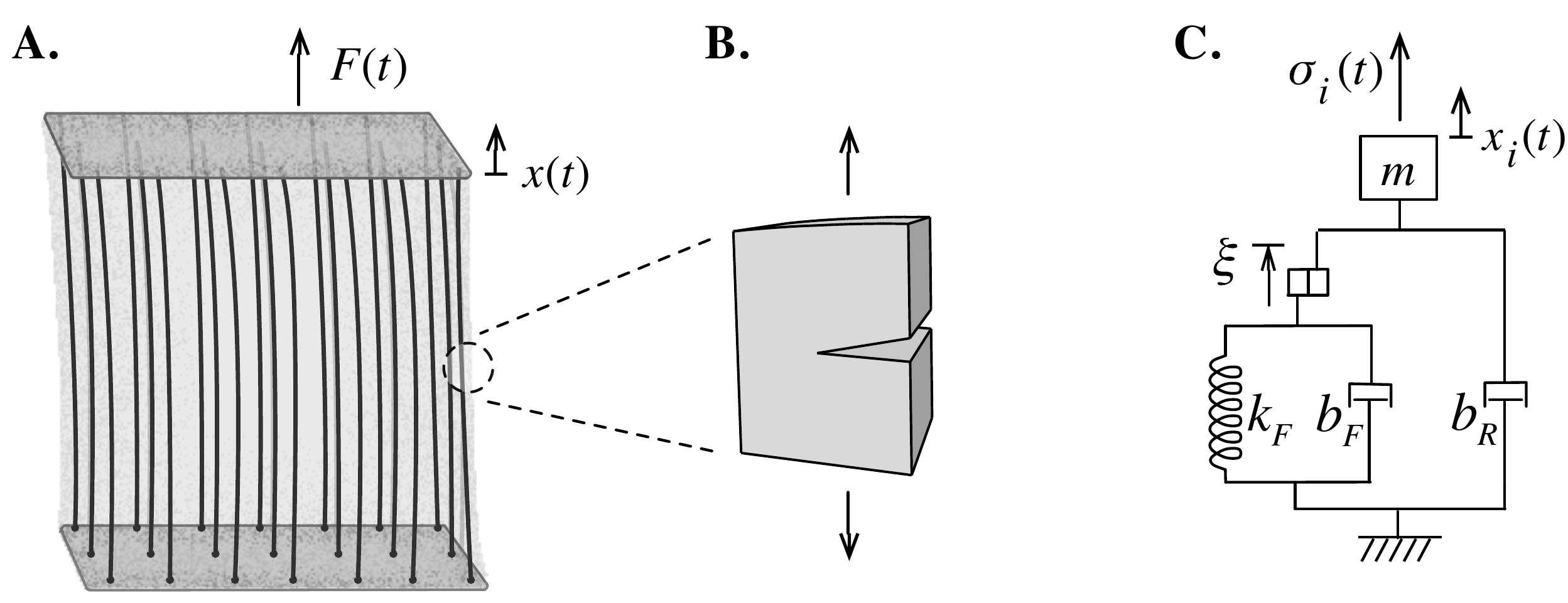}
\vspace{-2mm}
\caption[Fiber Bundle Model of fracture in a parallel array of $N$ brittle fibers.]{Fiber Bundle Model of fracture.  {\bf{A.}} A parallel array of $N$ brittle fibers is loaded  with external force $F(t)$.  {\bf{B.}} At the crack front, the weakest surviving fiber approaches failure due to local stress $\sigma_i$.  {\bf{C.}} A  mechanical analog in the form of a modified Kelvin-Vogt element.  A plastic unit breaks when the force on it is greater than a random threshold $\xi$, disconnecting spring $k_F$ and damper $b_F$.  Post-fracture relaxation is modeled via a  persistent damping factor, $b_R$.}
\label{fig:fbm}
\vspace{-3mm}
\end{center}
\end{figure}

Fiber Bundle Models (FBMs) are statistical lattice models of fracture capable of reproducing the most salient features of failure processes in heterogeneous materials \cite{alava2006statistical, pradhan2010failure}, including statistical strength distributions, stress fluctuations, reorganization accompanying failure, acoustic emissions, and accumulated damage, many of which are not well captured by standard continuum mechanics models of fracture.  They consist of coupled brittle elastic elements distributed over a spatial domain.  
FBMs can be expressed as lattices of $N$ parallel fibers each bearing a quantity $\sigma_i$, $i=1, 2, \ldots, N$, of the total mechanical force, $F$, on the bundle (Figure~\ref{fig:fbm}).   The strain $x_i(t)$ of the $i$th fiber is governed by a dynamical equation 
	\be
	F(t) = \sum_{i=1}^N \sigma_i(t) = \sum_{i=1}^N \left( \phi_i \sigma_i^F(t) + \sigma_i^R(t) \right)
	\label{eq:evolve}
	\ee
containing terms that represent the strain-dependent, per-fiber load $\sigma_i=\sigma_i(x_i,\dot{x}_i,\ddot{x}_i,\cdots)$ in terms of a part $\sigma^F_i$ born by intact fibers (for which the indicator variable $\phi_i=1$) and another, $\sigma^R_i$ by the surrounding matrix.  Only the latter persists after failure ($\phi_i=0$). A minimal micromechanical model capturing viscoelastic and plastic effects (Fig.~\ref{fig:fbm}, modified Kelvin-Vogt model) can be described by (\ref{eq:evolve}), with
	\be
	\sigma_i^F(t) = (b_F \dot{x} + k_F x), \ \sigma_i^R = (b_R \dot{x} + m \ddot{x}) 
	\ee
The dynamic response is parametrized by an effective (per-fiber) mass $m$, elastic constant $k_F$ and two damping constants $b_F$ and $b_R$ for the pre- and post-failure relaxation of the fiber.
The latter models creeping displacement in the matrix or sliding of fibers against it \cite{hidalgo2005slow}. Numerous variations on this micromechanical model are possible \cite{pradhan2010failure}, and can readily be accommodated in our treatment.  
 A fiber fractures when $x_i(t)$ exceeds a fiber-specific breaking threshold $\xi$.  The  thresholds are random variables, with $\xi \sim p(\xi)$.   After a fiber fractures, the load is redistributed among those that survive. 


We first assume equal load sharing (ELS) between intact fibers, so that the load on any intact fiber is $\sigma_i  = \sigma/N_F$, then discuss extensions to local load sharing (LLS). 
A fracture event decreases the number $N_F(t)$ of intact fibers at time $t$, increasing the load on surviving fibers, and cascading in further failures.  This continues until $x(t)<\xi^*$, where $\xi^*$ is the threshold of the weakest surviving fiber. When a critical value $F_c$ of the applied stress is reached, the bundle is incapable of supporting the redistributed load, and all remaining fibers break.  
The number $N_F$ of fibers surviving at a given load depends on the load history and random assignment of thresholds. Disorder is encoded in lattice initial conditions, and the subsequent evolution is deterministic.  Alternatively, one can regard the sequence of failure points as a random process whereby the fracture threshold $\xi$ jumps from one value to the next  at the time of fracture.

{\vspace{2mm}\hspace{-3.75mm}\textbf{Stochastic process formulation:}} Two key variables reflecting the instantaneous state of the model are the number $N_F$ of intact fibers and the breaking threshold $\xi^*$ of the weakest intact fiber.   
Upon failure, $\xi^*$ increases by a random amount $\Delta$ that is related to $p(\xi)$ and the number $N^*$ of preceding failures, where $N^* = N-N_F + 1$.   This can be interpreted as a stochastic jump process for a temporally fluctuating  threshold $\xi(t)$ that is defined to be equal at any instant $t$ to $\xi^*$, i.e., whose $j$th piecewise constant value $\xi(t) = \xi_j$ is reached at the instant $\xi(t-dt) = \xi_{j-1}$ is surpassed.  
The distribution of $k$th failures  can be described via  its order statistics \cite{kun2000damage}, but this obscures its character as a temporal process.  Instead, we propose to interpret the failure series as a sequential, monotonically increasing Markov process that reproduces the specified strength distribution $p(\xi)$.
To this end, we sequentially generate a series of monotonically increasing 
random variables $\xi_j$ that are distributed according to $p(\cdot)$, using the inverse transform sampling method \cite{Devroye1986}.  Let $u_j$ be independent samples of a random variable uniformly distributed  in $[0,1]$, for $j = 1, 2, \ldots, N,$ and set:
	\begin{eqnarray}
	\label{eq:randomsample0}
	s_0 &=& 0, \\
	\label{eq:randomsample}
	s_j &=& s_{j-1} + \left(1 - s_{j-1} \right) \left(1 - u_j^{1/\nfj}\right),  \\
	\xi_j &=& P^{-1}(s_j) \, .
	\label{eq:randomsample2}
	\end{eqnarray}
Here, $P(\cdot)$ is the CDF of the fiber strength distribution, $P^{-1}(\cdot)$ is its inverse function, and $\nfj=N-j+1$ is the number of surviving fibers prior to the $j$th failure.  The resulting sequence $\xi_j$ is equivalent to a set of $N$ independent samples from $p(\xi)$  sorted in increasing magnitude.  This algorithm reproduces the ensemble of samples from $p(\cdot)$ by sequentially sampling the conditional distributions
$p(\xi_j \, | \,  \xi_{j-1})
$.

Let $\xi(t) = \xi_j$ if $\xi_j$ is the failure strain of the weakest surviving fiber at time $t$.  
When a fracture event occurs,  $\xi(t)$ jumps in value and the number of surviving fibers, $N_F(t)$, decreases for each failed fiber.  This happens whenever the dynamic strain $x(t)$ exceeds $\xi(t)$.  
A fracture event at time $t$ is accompanied by the jump from $\xi$ to a new value $\xi'$ given by
 	\begin{eqnarray}
		s' &=& s + (1-s) \left(1-u^{1/N_F(t)}\right) \\
		\xi' &=& P^{-1}(s') .
		\label{eq:jump1}
	\end{eqnarray}
This  
yields a simple iterative expression for $\xi'$ in terms of $\xi$ and $N_F(t)$:	
\begin{equation}
		\xi'  = P^{-1}\biggl(P(\xi) + \bigl(1-P(\xi)\bigr) \Bigl( 1 - u^{1/N_F(t)} \Bigr) \biggr) .
		\label{eq:jump2}
	\end{equation}
The size of a jump in $\xi$ at time $t$ depends on the state of the co-evolving random process $N_F(t)$ and on the value of $\xi(t)$, while the time at which it occurs depends on the values of $\xi(t)$ and the strain $x(t)$.  

{\it Local Model of Continuous Damage:} 
This model can be viewed as capturing a domain of brittle elements by a representative fiber 
undergoing repeated fracture displacements of size $S(t_i) = \xi(t_i+dt) - \xi(t_i)$ at times $t_i$, $i= 1, 2, \ldots, N$.  
In this light, the foregoing can be interpreted as an effective model of accumulated damage, as in the Continuous Damage Model of Kun et al.~\cite{kun2000damage}.  Equation (\ref{eq:jump2}) shows how a continuous damage description can be derived from a distributed micromechanical model of failure at a smaller length scale - one that is ``integrated out'' to yield the multiple-failure process $\xi(t)$.  


{\vspace{2mm}\hspace{-3.75mm}\textbf{Mean field approximation:}} Our global strain threshold $\xi(t)$ depends on the level of damage at the time of fracture (represented by $1-N_F$), itself a random value that depends on the history of  the sample.  
Under ELS, its expected value is $\bar{N}_F = N (1 - P(x))$, where $x$ is the maximum strain achieved during loading. 
The instantaneous strain $x(t)$ depends on the stochastic evolution of damage in the lattice.  However, the dynamics will tend to average the fluctuating stresses.  This suggests that we may average over fluctuations to approximate the nonlinear strain evolution deterministically, with random effects entirely captured by the variable $\xi$.  This is simply achieved by replacing $N_F(t)$, where it appears in the threshold and evolution equations (\ref{eq:jump2}) and (\ref{eq:evolve}), by the expected survival number $\bar{N}_F(t)$ given the load history.  $\bar{N}_F$ represents the mean damage that would be expected for an ensemble of instances of the model subjected to the given load history.

The expected survival number $\bar{N}_F(t)$ depends on the strain via
\be
	\bar{N}_F(t) = N \left( 1 - P(x^*(t))\right), \ \ \ x^*(t) = \max_{t'<t}x(t') .
\ee
Under monotonically increasing loading, this equals $\bar{N}_F(t) = N(1-P(x(t)))$.  Upon replacing $N_F$ by $\bar{N}_F$ 
one can factorize the model into a deterministic part governing the nonlinear stress-strain response, 
\be
	F(t) = \bar{N}_F(t) \sigma^F(t) + N \sigma^R(t) \\
	\label{eq:deterministic}
\ee
and a stochastic process describing the stress fluctuations:
\bea
	\label{eq:stochastic}
	 &&  \xi'  = P^{-1}\biggl(P(\xi) + \bigl(1-P(\xi)\bigr) \Bigl( 1 - u^{1/\bar{N}_F} \Bigr) \biggr) \\
	&& \mathrm{while} \ x(t) > \xi \notag .
\eea
We refer to the original FBM model as $M$ and the approximation obtained through this ``mean damage'' replacement as $\bM$.  
The latter takes on explicit form only after a fiber strength distribution $p(\xi)$ is specified. 

{\vspace{2mm}\hspace{-3.75mm}\textbf{Uniform distribution:}} 
When $p(x)$ is uniform on $[0,1]$, assuming monotonic loading, $x^*(t) = x(t)$, hence $\bar{N}_F(t) = N (1-x(t))$. Assuming, for illustration, a modified Kelvin-Vogt micromechanical model as in (2), the homogenized nonlinear stress-strain response becomes:
\be
	F(t) = N (1-x) \, (b_F \dot{x} + k_F x) + N (b_R \dot{x} + m \ddot{x})
\ee
while the increased threshold $\xi'$ is sampled as: 
\be
	\xi'  = \xi + (\xi-1)^2 (u^{1/(N(1-\xi))} - 1) ,
\ee
where $u$ is a sample of a random variable uniformly distributed in $[0,1]$.


\begin{figure}[h]
\begin{center}
\includegraphics[width=8.cm]{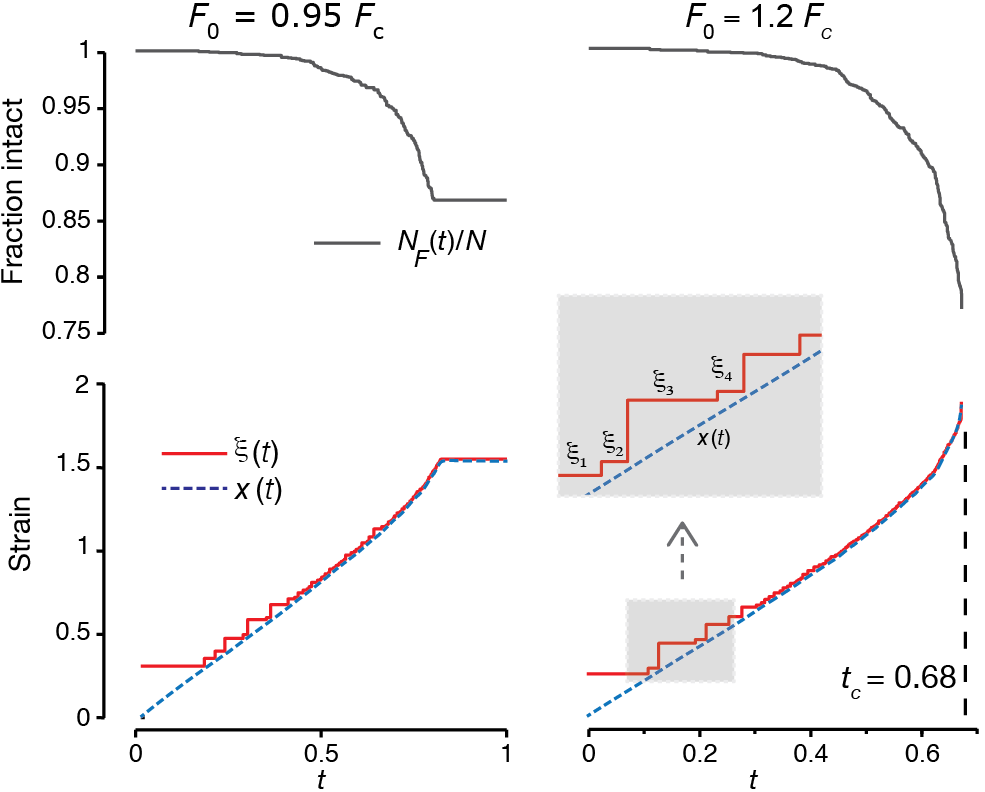}
\vspace{-.25cm}
\caption[Typical stress-controlled loading responses of the Fiber Bundle Model.]{Two instances of  responses to a linear ramp load, $F(t) = F_0 r(t)$, where $r(t)=t/\tau_0$ for $0 \leq t \leq \tau_0 = 0.8$.  $N= 2000$ fibers, and Weibull-distributed thresholds (parameters $k=4$ and $\lambda=2.5$).  Top row: Survival fraction $N_F(t)/N$ vs $t$. Maximum stress, $F_0$, is shown  relative to critical stress, $F_c\approx 2800$.  Right side: $F_0 > F_c$.  Failure occurs at $t = 0.68$.  Bottom row: Threshold $\xi(t)$ and  strain $x(t)$ vs $t$. } 
\label{fig:results1}
\end{center}
\end{figure}

{\vspace{2mm}\hspace{-3.75mm}\textbf{Weibull distribution:}} This has been found to be a good empirical statistical distribution for solid strength in materials science. The Weibull CDF is given by
	\be
		P(\xi) = 1 - \exp(\xi/\lambda)^k \Theta(\xi)
	\ee
where $\lambda$ and $k$ are scale and shape parameters, and $\Theta(\cdot)$ is the Heaviside step function, with $\Theta(\xi) = 0$ for $\xi < 0$ and $\Theta(\xi) = 1$ for $\xi > 0$.  For this choice of distribution, the mean damage model can be written as
\be
	\sigma(t) = N \exp(- x / \lambda)^k \, (b_F \dot{x} + k_F x) + N (b_R \dot{x} + m \ddot{x})
	\label{eq:deterministic2}
\ee
with the jump process for the failure threshold $\xi$ given by
\[
	\xi' =  \lambda \, \left(\frac{\xi}{\lambda}\right)^k  \log\left(
		1 - e^{\left(-\xi/\lambda\right)^k} (1-u^{1/(N\exp(-\xi/\lambda)^k)})  \right)^{1/k} 
\]

\begin{figure}[]
\begin{center}
\includegraphics[width=8.45cm]{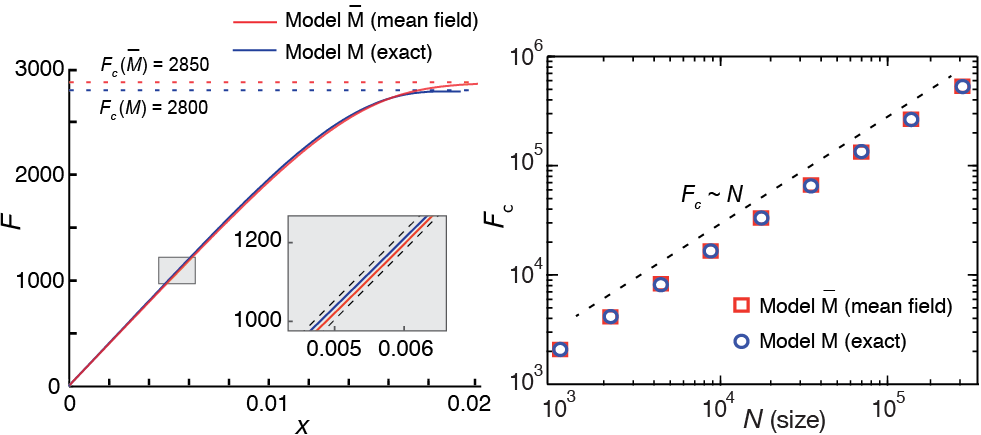}  
\vspace{-.25cm}
\caption[The constitutive stress-strain relationship $\sigma(x)$ for models $M$, $\tm$, and $\bM$.]{Empirical constitutive law $\sigma(x)$ for  $M$ and $\bM$.  Simulations obtained under slow ramp loading conditions, at supra-critical values of the maximum stress, $F>F_c$,  computed from 200 uniformly distributed values of the terminal load, using  simulations with 2000 fibers.    Dashed blue line: critical stress values of $F_c (M)= 2800, F_c(\bM) = 2850$.  Dashed red lines:  95\% confidence intervals.}
\label{fig:results2}
\end{center}
\end{figure}

{\vspace{2mm}\hspace{-3.75mm}\bf Constitutive behavior and fluctuations:} 
The value of the critical stress $F_c$ and distributions  of fluctuations as failure approaches are known to depend  weakly on the precise distribution of fiber strengths \cite{alava2006statistical,pradhan2010failure}.  
Specifically, the constitutive evolution of $\bM$ differs from that of $M$ due to fluctuations in the survival number $N_F$ about its mean.  This can be regarded as a source of high-frequency noise that should approximately integrate to zero, so we reasoned that even if these fluctuations are significant, the model $\bM$ would yield similar behavior to $M$. To test this, we numerically simulated both systems to obtain 
stress-strain constitutive relations, critical load, and failure distributions under stress-controlled loading, using a Weibull strength distribution.  When $N$ is large, due to the frequent fracture events, the equations for model $M$ behave like a stiff ODE, so we employed a fine-grained variable time step implicit ODE solver with both. 
Simulation runs are qualitatively indistinguishable for both models, see Figure~\ref{fig:results1} (samples of model $M$).     



Stress-strain relationships and critical load estimates are compared for both models in Figure \ref{fig:results2}.  
Qualitatively and quantitatively, both are nearly identical, with an error of less than $2\%$ in the critical load $F_c$ for all bundle sizes 
examined (size $N=1000$ to $256000$). 

\begin{figure}[t] 
\begin{center}
\includegraphics[width=8.0cm]{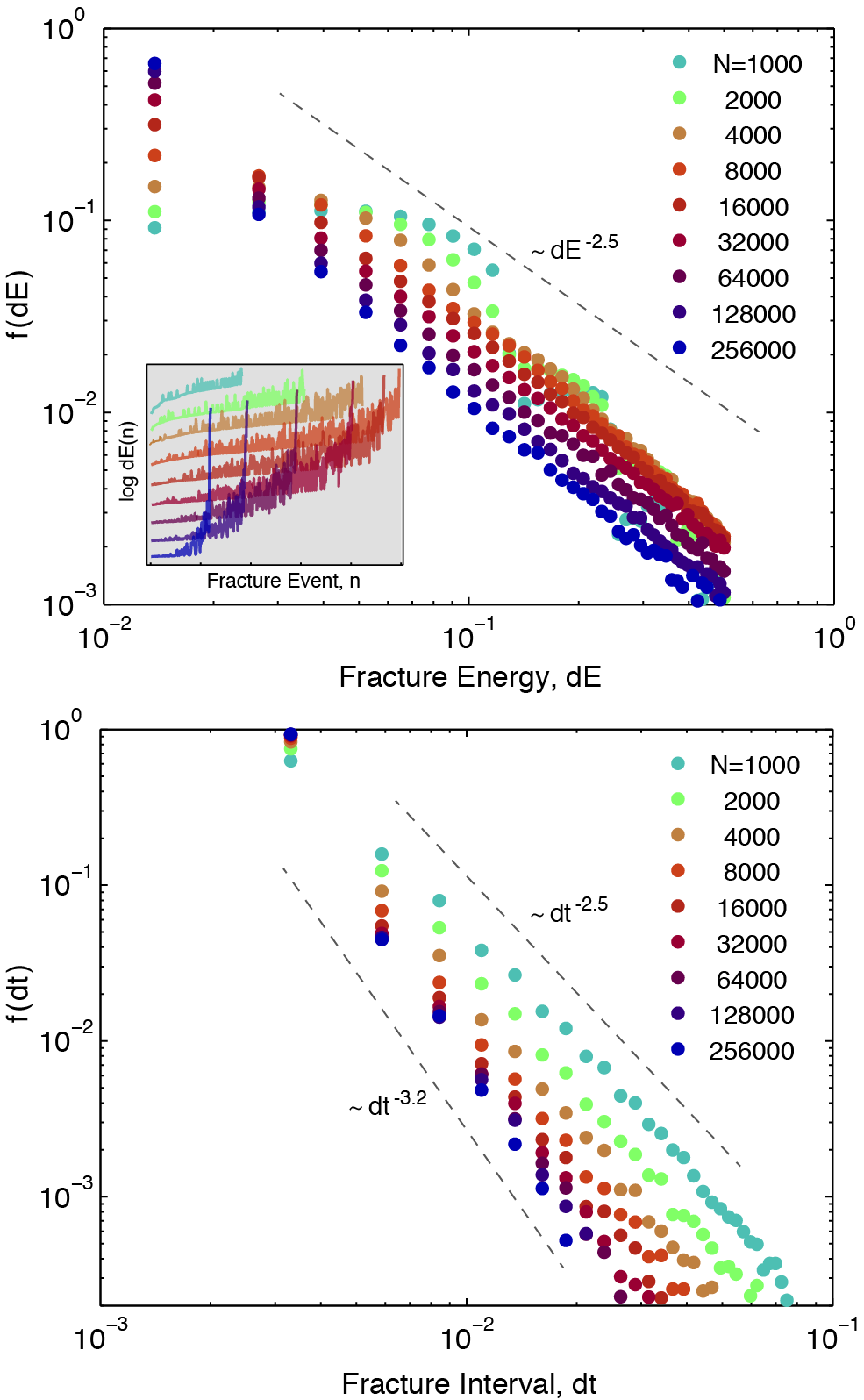} 
\vspace{-.25cm}
\caption[Empirical distribution of the energy $dE$ of fracture fluctuations.]{Top: Empirical distribution of fracture energy $dE$, showing approximate power law scaling $\sim dE^{-2.5}$ consistent with published results \cite{alava2006statistical,pradhan2010failure}.  Finite-size effects can be observed for small $N$.  Distributions were estimated from 1000 simulations of bundles of each size $N$ subjected to ramp loading.  Top, Inset: Energy burst size vs burst event during a representative trial at each size. Bottom: The distribution of inter-event times $dt$ also evidences power-law scaling.}
\label{fig:results3b}
\vspace{-4mm}
\end{center}
\end{figure}

Figure  \ref{fig:results3b} examines the empirical distributions of failure event time intervals $dt$ and of energy fluctuations $dE$ for model $\bar{M}$ (results for $M$ are effectively identical).  The fracture of a fiber at strain  $x_i$ releases elastic energy $dE = k_F x_i^2$.  Since our treatment is dynamic, to estimate the distribution of energy fluctuations, energy released by all events within each time window of duration 0.005 was integrated.  Above small values of $N$, where finite-size effects are apparent, The results exhibit approximate power law scaling, consistent with expected fracture behavior approaching critical failure  \cite{pradhan2010failure}.  
\\[2mm]

{\bf\hspace{-4.75mm} Local load sharing:}
The ELS assumption is simplifying, but unphysical for large samples \cite{batrouni2002heterogeneous, herrmann1990statistical,hansen1994burst,kun2003creep,hidalgo2002fracture}, as the per-fiber stress $\sigma_i(t)$ is differently affected by remote fiber failures. This can be quantified through a factor $A_i$ that enhances the stress of an intact fiber after a failure, such that $\sigma_i \rightarrow A_i \sigma_i(t) = \phi_i \sigma^F_i(t) + \sigma^R_i(t)$, with
\[
    A_i = Z^{-1} \sum_{j=1}^N (1-\phi_j) F_{ij}, \ \ Z = \sum_{i=1}^N \phi_i \sum_{j=1}^N (1-\phi_j) F_{ij} 
\]
The weight $F_{ij} \sim r_{ij}^{-\gamma}$ models the reduction of load transfer with distance $r$, and the failure indicator variable  
$\phi_i$ captures the spatial fracture pattern. We briefly describe how to accommodate stress enhancement in our model.  Assuming a uniform spatial distribution of fibers, one can compute a probability distribution $p(A)$ of load transfer factors $A$, with the result $p(A) \propto \gamma^{-1} A^{(\gamma+2)/\gamma}$ \cite{lehmann2010breakdown}. We can capture multi-fracture stress enhancement through a factor $\hat{A} = A_1 A_2 \cdots A_{N-N_F}$, where $A_k$ are independent samples of $p(A)$ for each failure.  The stress-enhanced version of the homogenized dynamical equation (\ref{eq:deterministic}) becomes $\hat{A} F(t) = \bar{N}_F(t) \sigma^F(t) + N \sigma^R(t)$.
\\[2mm]

The FBM formulation presented here describes random failure evolution through a stochastic jump process governing failure thresholds, coupled to a mean-field  approximation to damage accumulation. This factorization was achieved without impairing
accuracy. The method can accommodate a wide range of micromechanical
models for individual fibers, including non-negligible dynamics or nonlinearity.  
The result is efficient enough
to allow simulation of stress fluctuations in large bundles in real time, 
which could further aid applications in scientific simulation and visualization; See supplementary material [URL] for multimedia documentation.


%

\bibliography{fbm}

\end{document}